\documentstyle[epsfig]{aa}

\newcommand{\Msolar}{\mbox{\,$\rm M_{\odot}$}}        

\begin{document}
\title
{Masses, Accretion Rates and Inclinations of AGNs}
\author
{W.Bian \inst{1,2} \and Y.Zhao\inst{1}} \institute{National
Astronomical Observatories, Chinese Academy of
 Sciences, Beijing 100012, China
\and Department of Physics, Nanjing Normal University, Nanjing
210097, China}

\offprints{W.Bian,
\\
\email{whbian@njnu.edu.cn}}

\date{Received ...; accepted ...}

\titlerunning{Masses, Accretion Rates and Inclinations}
\authorrunning{W.Bian \& Y.Zhao}
\abstract{ We assume that the gravitational instability of
standard thin accretion disks leads to the Broad Line Regions
(BLRs), the B band luminosity comes from standard thin disk and
the motion of BLRs is virial. The central black hole masses, the
accretion rates and the disk inclinations to the line of sight for
17 Seyfert 1 galaxies and 17 Palomar-Green (PG) quasars have been
calculated. Our results are sensitive to $\alpha$ parameter of the
standard $\alpha$ disk. With the same values of $\alpha$
($\alpha=1$), calculated central black hole masses for 17 Seyfert
1 galaxies are consistent with that from Kaspi et al. (2000) while
that for 17 PG quasars are larger than that from Kaspi et al.
(2000) by almost 2 orders of magnitude. Inclinations of 17 Seyfert
1 galaxies are about 6 times larger than that of 17 PG quasars.
These inclinations, with a mean value of $32^{o}$ for 17 Seyfert 1
galaxies that agrees well with the result obtained by fitting the
iron $K\alpha$ lines of Seyfert 1 galaxies observed with ASCA
(Nandra et al. 1997) and the result obtained by Wu \& Han (2001),
provide further support for the orientation-dependent unification
scheme of active galactic nuclei. There is a relation between the
FWHM of H$\beta$ and the inclination, namely the inclination is
smaller in AGNs with smaller FWHM of H$\beta$. The effect of
inclinations in narrow line Seyfert 1 galaxies (NLS1s) should be
considered when one studies the physics of NLS1s. The need of
higher value of $\alpha$ for PG quasars maybe shows that our model
is not suitable for PG quasars if we think inclinations of PG
quasars are in the inclination levels of Seyfert 1 galaxies. With
our model, we also show that the size of BLRs relates not only to
the luminosity, but also to the accretion rates. More knowledge of
BLRs dynamics, accretion disks and optical luminosity are needed
to improve the determinations of black hole masses, accretion
rates and inclinations in AGNs.
}

\maketitle

\section{Introduction}

\label{intro}

Broad emission lines are one of the dominant features of many
Active Galactic Nuclei (AGNs) spectrum. Broad Line Regions (BLRs)
play a particularly important role in our understanding of AGNs by
virtue of its proximity to the central source. With reverberation
mapping technique the sizes of the BLRs can be obtained through
the study of correlated variations of the lines and continuum
fluxes (Peterson 1993). The BLR sizes for 17 Seyfert 1 galaxies
(Wandel et al. 1999) and for 17 PG quasars (Kaspi et al. 2000)
have been recently obtained. Assuming the Keplerian motions and
the random orbits of the BLRs, the central black hole masses also
have been obtained. In order to underline the physics of AGNs,
these masses have been used to check the relations with other
parameters of AGNs, such as the radio power (Ho 2002), the X-ray
excess variance (Lu \& Yu 2001), the velocity dispersions of their
host galaxies (Ferrarese et al. 2001; Gebhardt et al. 2000b).

Based on the unified model of AGNs, the wide variety of AGN
phenomena we see is due to a combination of real differences in a
small number of physical parameters (e.g. luminosity) coupled with
apparent differences which are due to observer-dependent
parameters (e.g. orientation). AGNs with wide emission lines from
BLRs are oriented at a preferred angle from which BLRs is visible.
No edge-on disks in AGNs with wide emission lines would be seen
(Urry \& Padovani 1995). Therefore, inclinations of AGNs with wide
emission lines are generally expected to be small although further
evidence is obviously needed. It is important to test it with
inclinations of AGNs. Random mean angle is often assumed in
calculated black hole masses with the reverberation mapping
method. Some authors (McLure \& Dunlop 2001; Wu \& Han 2001) have
shown it is necessary to consider the effects of the inclination.

Though it is powerful for size determination, currently
reverberation mapping method does not provide the velocity field
of the BLRs, and can not distinguish between radial and rotational
motions. Because there is no consensus about the dynamics of the
BLRs, there has been little progress in understanding the physics
of this region. Recently some authors (Collin \& Hure 2001)
suggest the gravitationally unstable disc is the source which
releases BLRs in the medium.

In this paper, we also assume the gravitational instability leads
to the formation of BLRs. With the analysis formulae of standard
thin accretion disks and the observational sizes of the BLRs, we
calculate the central masses, accretion rates and inclinations for
34 AGNs. We want to give the clues to the origin of BLRs, the
values of three parameters of accretion disks and the inclination
effect in mass determination in AGNs. The paper is structured as
follows. In Sect. 2, we describe our model and available data. In
sect. 3, we give our calculational results. Our discussion and
conclusions are presented in the last section.

\section{Calculation}

\subsection{Formulae}

First, we consider gravitational potential energy at R away from
the center of the black hole is released at the rate
$GM\dot{M}/R$, where $M$ is the central black hole mass and
$\dot{M}$ is the accretion rate; from the virial theorem, half of
this goes into heating the gas, and the other half is radiated
away. Thus the disk temperature at R is
$T=(\frac{3GM\dot{M}}{8\pi\sigma R^{3}}(1-(R/R_{s})^{1/2}))^{1/4}$
(Frank et al. 1992), where G is the gravitational constant,
$\sigma$ is Stefan-Boltzmann constant,
 $R_{s}=2GM/c^{2}$ is the Schwarzchild radius of the black hole
with the mass of M. For $R>>R_{s}$, this can be simplified,
$T=(\frac{3GM\dot{M}}{8\pi\sigma
R_{s}^{3}})^{1/4}(R/R_{s})^{-3/4}=T_{\star}(R/R_{s})^{-3/4}$. As a
first approximation we can assume the accretion disk radiates
locally like a blackbody and the specific luminosity is
$L_{\nu}=2.4\times10^{-18}R_{s}^{2}cosiT_{\star}^{8/3}\nu^{1/3}
~~\rm {ergs ~s^{-1} Hz^{-1}}$ (Peterson 1997). We obtained the B
band luminosity,
$L^{B}=0.5312\times10^{44}\dot{M}^{2/3}M^{2/3}cosi ~~\rm {ergs~
s^{-1}} $, where i is the inclination of the disk to the line of
sight and the frequency is between 3900$\AA$ and 4900$\AA$. We can
rewrite it in terms of $R_{14}=R/(10^{14}cm)$,
$M_{8}=M/(10^{8}\Msolar)$, $L^{B}_{9}=L^{B}/(10^{9}L\sun)$ and
$\dot{M}_{26}= \dot{M}/(10 ^{26} \rm{g~s^{-1}})$,

\begin{equation}
L^{B}_{9}=13.8\dot{M}_{26}^{2/3}M_{8}^{2/3}cosi.
\end{equation}

Second, we assume that the gravitational instability of the disk
at large radii leads to the formation of BLRs. The criterion is
$Q=\frac{\Omega^{2}}{\pi G\rho}\leq 1$ (Golreich \& Lynden-Bell
 1965) , where $\Omega$ is the Keplerian angular velocity
$(GM/R^{3})^{1/2}$ and $\rho$ is the local mass density. We
adopted the solutions of $\rho$ for the standard thin disk from
Shakura \& Sunyaev (1973),
$\rho=3.1\times10^{-5}\alpha^{-7/10}\dot{M}_{26}^{11/20}
M_{8}^{5/8}R_{14}^{-15/8}f^{11/5} \rm{~~g~cm^{-3}}$, where
$f=(1-(R/R_{s})^{1/2})^{1/4}$, $\alpha$ is the parameter of the
standard $\alpha$ disk. We adopted $f=1$. The value of $\alpha$ in
AGNs is often between 0 and 1. We obtained the size of the BLRs,
\begin{equation}
R_{14}=880\alpha^{28/45}Q^{-8/9}\dot{M}_{26}^{-22/45}M_{8}^{1/3}.
\end{equation}

Third, assuming the Keplerian velocity of the BLRs, then the
so-called virial mass estimated for the central black hole is
given by $M=\frac{V_{FWHM}^{2}}{4(sini^{2}+A^{2})}RG^{-1}$ (McLure
\& Dunlop 2001), where $V_{FWHM}$ is the FWHM of the H$\beta$
emission line, i is the inclination, A is the ratio of the random
isotropic characteristic velocity to the Keplerian velocity of the
disk and A is omitted here (Wu \& Han 2001).
 Then we can get,
\begin{equation}
V_{3}=3.89\alpha^{-14/45}\dot{M}_{26}^{11/45}M_{8}^{1/3}sini.
\end{equation}
where $V_{3}=V_{FWHM}/(1000 \rm{km~s^{-1}})$.

\subsection{Data}

The absolute B band magnitudes are adopted from Veron-Cetty et al.
(2001), which are calculated by adopting an average optical index
of -0.5 and accounting for Galactic redding and K-correction.
Giveon et al. (1999) have calculated the median absolute magnitude
for the PG sample that Kaspi et al. (2000) have used. The values
of absolute B band magnitude from Giveon et al. (1999) are often
different from the values we adopted from Veron-Cetty et al.
(2001). The optical variability is a common property for AGNs
(Giveon et al. 1999). The B magnitude of Seyfert 1 galaxies might
be affected from their host galaxies. Therefore we adopt one B
band magnitude uncertainty when we calculate the uncertainty of
our calculated results. The sizes of the BLRs for 17 Seyfert 1
galaxies and 17 PG quasars are adopted from Kaspi et al. (2000).
The central black hole masses calculated from reverberation
mapping are also adopted from Kaspi et al. (2000). The properties
of 34 AGNs are listed in Table 1. Column 2,3,4 are respectively
absolute B band magnitudes, sizes of BLRs and FWHMs of $H\beta$.

\section{Results}

Using Eq. (1), Eq. (2), Eq. (3), we can calculate the central
black hole masses ($M$) , the accretion rate ($\dot{M}$) and the
inclination ($i$) knowing absolute B band luminosity ($L^{B}$),
sizes of the BLRs  ($R$) and FWHMs of H$\beta$ ($V_{FWHM}$).
Errors of FWHMs of H$\beta$ are not included in our calculation.
$\alpha$ is 1 in all the calculations and the uncertainties of our
calculated results are estimated by considering $1/4 \leq Q \leq
4$ (Golreich \& Lynden-Bell 1965), errors of sizes of BLRs and one
absolute B band magnitude uncertainty except in situations where
we declare otherwise. We use Q=0.25, the lower limits of BLRs
sizes and the upper limits of the absolute B magnitude to give one
directional limits of our results. We calculate the another
directional limits of our results using Q=4, the upper limits of
BLRs sizes and the lower limits of the absolute B magnitude. We
present the accretion rate in units of Eddington accretion rate,
$\dot{m}=\dot{M}_{26}1.578/(3.88M_{8})$ (Laor \& Netzer 1989). The
results are shown in Table 1. Column (6) is our calculated BH
masses. Column (7) is inclinations of the disk to the line of
sight. Column (8) and Column (9) are the accretion rates and the
accretion rate in unites of Eddington accretion rate,
respectively. Column (10) lists the the radio loudness (Nelson
2001). Column (11) is the available nuclear radio loudness (Ho \&
Peng 2001). Column (12) is inclinations for the same AGNs from Wu
\& Han (2001).

We also use the bootstrap method, which uses the actual data and
their errors to generate many data following a gaussian
distribution, to calculate uncertainties of our results. Based on
the measured values and their errors of some parameters (such as
BLRs sizes, FWHMs of $H\beta$ line, B magnitude), independent
20000  gaussian distribution data points of these parameters are
generated, which are used to drive 20000 values of our results. We
can get the values and uncertainties of our results using gaussian
distribution to fit 20000 results. The values and uncertainties of
our results from the bootstrap method are almost the same as our
results listed in Table 1.
\begin{table*}
\begin{center}
\begin{tiny}
\begin{tabular}{llllllllllll}
\hline \hline

 Name      &$M_{B}$ & $R_{BLR}$ & $V_{H\beta}$&  $log_{10}M_{rm}$ &  $log_{10}M_{cal}$   & $i$  &  $log_{10}\dot{M}$  & $\dot{m}$  &  $R_{ro}$      & $R_{nucl}$  & $i_{disp}$\\
            &(Mag)  &(lt~ days) &(1000km/s)&  $(\Msolar)$      & $(\Msolar)$    &   (deg)  & $(\Msolar/yr)$  &    &                    &         &(deg)    \\
(1)&(2)&(3)&(4)&(5)&(6)&(7)&(8)&(9)&(10)&(11)&(12)\\
 \hline
 3C120      & -20.8 & $42^{+27}_{-20}     $& $1.91\pm0.12$   & $7.36_{-0.28}^{+0.22}$ & $8.19 ^{+1.26}_{-1.32} $  & $12.6_{- 5.7}^{+10.5}$& $0.14^{+0.76}_{-0.67}$  &$0.228 _{-0.2}    ^{+27.9}$     &  16.8  & --  & $22.0^{+9.3}_{-7.7}$    \\
 3C390.3    & -21.6 & $22.9^{+6.3}_{-8.0} $& $10  \pm0.8 $   & $8.53_{-0.21}^{+0.12}$ & $8.30 ^{+1.02}_{-0.94} $  & $48.0_{-19.7}^{+24.0}$& $0.76^{+0.84}_{-0.61}$  &$0.728 _{-0.7}    ^{+42.6}$     &  198   & -- & --    \\
 Akn120     & -22.2 & $37.4^{+5.1}_{-6.3} $& $5.8 \pm0.48$   & $8.26_{-0.12}^{+0.08}$ & $8.65 ^{+1.06}_{-1.04} $  & $21.7_{- 9.2}^{+15.6}$& $0.56^{+0.55}_{-0.49}$  &$0.207 _{-0.2}    ^{+7.9} $      &  0.25  & --  & --   \\
 F9         & -23.0 & $16.3^{+3.3}_{-7.6} $& $5.78\pm0.65$   & $7.9 _{-0.31}^{+0.11}$ & $8.49 ^{+1.09}_{-1.09} $  & $17.1_{- 7.3}^{+12.7}$& $1.18^{+0.55}_{-0.51}$  &$1.277 _{-1.2}    ^{+54.8}$     &  0 & --    &-- \\
 IC4329A$^{a}$    & -20.1 & $ 1.4^{+3.3}_{-2.9} $& $5.05\pm2.07$   & $6.7 _{-3.70}^{+0.56}$ & $6.38 ^{+1.47}_{-1.82} $  & $58.0_{-25.8}^{+25.1}$& $1.93^{+2.19}_{-1.17}$  &$903.80_{-901.7}  ^{+9.4E6}$&  10.2 & --  &--    \\
 Mrk79      & -20.9 & $17.7^{+4.8}_{-8.4} $& $4.47\pm0.85$   & $7.72_{-0.34}^{+0.14}$ & $7.82 ^{+1.09}_{-1.20} $  & $30.6_{-13.1}^{+22.5}$& $0.66^{+0.83}_{-0.56}$  &$1.76  _{-1.7}    ^{+185.8}$     &  0.8   & -- &$58.5^{+21.7}_{-27.9}$     \\
 Mrk110     & -20.6 & $18.8^{+6.3}_{-6.6} $& $1.43\pm0.12$   & $6.75_{-0.20}^{+0.13}$ & $7.70 ^{+1.16}_{-1.22} $  & $11.2_{- 4.9}^{ +8.9}$& $0.52^{+0.65}_{-0.56}$  &$1.70  _{-1.7}    ^{+123.6}$     &  1.6   & -- & $37.4^{+9.2}_{-9.5}$   \\
 Mrk335     & -21.7 & $16.4^{+5.1}_{-3.2} $& $1.62\pm0.12$   & $6.8 _{-0.14}^{+0.14}$ & $8.01 ^{+1.15}_{-1.11} $  & $8.2 _{- 3.6}^{ +6.2}$& $0.85^{+0.53}_{-0.55}$  &$1.79  _{-1.7}    ^{+76.5}$      &  0.2   & 0.62 &--  \\
 Mrk509     & -23.3 & $76.7^{+6.3}_{-6.0} $& $2.27\pm0.12$   & $7.76_{-0.05}^{+0.05}$ & $9.40 ^{+1.05}_{-1.04} $  & $5.0 _{- 2.1}^{ +3.7}$& $0.43^{+0.45}_{-0.45}$  &$0.028 _{-0.02}   ^{+0.8} $     &  0.4   & --    &-- \\
 Mrk590     & -21.6 & $20.0^{+4.4}_{-2.9} $& $2.47\pm0.12$   & $7.25_{-0.09}^{+0.10}$ & $8.09 ^{+1.11}_{-1.07} $  & $12.7_{- 5.5}^{ +9.4}$& $0.73^{+0.50}_{-0.52}$  &$1.13  _{-1.1}    ^{+41.1}$      &  0.5   & 1.62 &$17.8^{+6.1}_{-5.9}$  \\
 Mrk817     & -22.3 & $15.0^{+4.2}_{-3.4} $& $4.49\pm0.18$   & $7.64_{-0.12}^{+0.11}$ & $8.19 ^{+1.12}_{-1.10} $  & $17.9_{- 7.7}^{+13.3}$& $1.06^{+0.57}_{-0.55}$  &$1.88  _{-1.8}    ^{+86.6}$      &  0.8   & 1.21  &$41.6^{+8.5}_{-7.5}$ \\
 NGC3227$^{b}$    & -18.7 & $10.9^{+5.6}_{-10.9}$& $4.92\pm0.49$   & $7.59_{-4.60}^{+0.19}$ & $7.16 ^{+0.94}_{-2.60} $  & $71.1_{-24.1}^{+17.1}$& $0.63^{+3.49}_{-0.82}$  &$7.70  _{-7.6}    ^{+9.4E6}$     &  4.7   & 1.12  &$37.5^{+17.5}_{-25.4}$ \\
 NGC3783    & -19.7 & $4.5 ^{+3.6}_{-3.1} $& $3.79\pm1.16$   & $6.97_{-0.97}^{+0.30}$ & $6.76 ^{+1.21}_{-1.27} $  & $47.5_{-20.5}^{+26.7}$& $1.15^{+1.26}_{-0.79}$  &$62.8  _{-62.2}   ^{+21545.3}$ &  0.6   & -- & --    \\
 NGC4051    & -16.8 & $6.5 ^{+6.6}_{-4.1} $& $1.17\pm0.06$   & $6.11_{-0.41}^{+0.30}$ & $5.91 ^{+1.28}_{-1.19} $  & $46.4_{-20.3}^{+27.4}$& $0.25^{+1.17}_{-0.85}$  &$55.11 _{-54.7}   ^{+12478.6}$ &  6 & 0.87 &$19.5^{+10.4}_{-6.6}$  \\
 NGC4151    & -18.7 & $3.0 ^{+1.8}_{-1.4} $& $5.23\pm0.92$   & $7.18_{-0.38}^{+0.23}$ & $6.62 ^{+0.89}_{-0.83} $  & $77.2_{-22.1}^{ +9.5}$& $1.42^{+1.09}_{-0.90}$  &$159.5 _{-157.0}  ^{+12998.9}$ &  5.6   & 0.49&$60.0^{+30.0}_{-30.6}$   \\
 NGC5548    & -20.7 & $21.2^{+2.4}_{-0.7} $& $6.3 \pm0.4 $   & $8.09_{-0.07}^{+0.07}$ & $7.92 ^{+0.98}_{-0.79} $  & $44.7_{-18.1}^{+22.8}$ & $0.56^{+0.59}_{-0.52}$  &$1.13  _{-1.1}    ^{+25.7}$   &  1.5   & 1.24   &$43.7^{+7.6}_{-6.9}$\\
 NGC7469    & -21.6 & $4.9 ^{+0.6}_{-1.1} $& $3   \pm1.58$   & $6.81_{-3.81}^{+0.30}$ & $7.35 ^{+1.06}_{-1.1 } $  & $18.0_{-7.6}^{+13.4} $ & $1.48^{+0.57}_{-0.48}$  &$34.23 _{-33.2}   ^{+1568.6 }$&  3 & 3.2   &-- \\
 PG0026     & -24.0 & $113 ^{+18}_{-21}   $& $2.1  \pm0.14  $& $7.73_{-0.10}^{+0.07}$ & $9.85 ^{+1.09}_{-1.11} $  & $3.4 _{-1.4}^{+ 2.6}$ & $0.39^{+0.51}_{-0.49}$  &$0.009 _{-0.009}  ^{+0.4}$   & 1.6     & --&--          \\
 PG0052     & -24.5 & $134 ^{+31}_{-23}   $& $3.99 \pm0.24  $& $8.34_{-0.12}^{+0.11}$ & $10.12^{+1.12}_{-1.10} $  & $5.1 _{-2.2}^{+ 3.9}$ & $0.42^{+0.51}_{-0.52}$  &$0.0052_{-0.005}  ^{+0.2}$   & 0.3     & -- &--         \\
 PG0804     & -23.9 & $156 ^{+15}_{-13}   $& $2.98 \pm0.051 $& $8.28_{-0.04}^{+0.04}$ & $9.98 ^{+1.06}_{-1.05} $  & $4.8 _{-2.0}^{+ 3.5}$ & $0.20^{+0.46}_{-0.46}$  &$0.0042_{-0.004}  ^{+0.1}$  & 0.2     & --  &--        \\
 PG0844     & -23.1 & $24.2^{+10}_{-9.1}  $& $2.73 \pm0.12  $& $7.33_{-0.02}^{+0.02}$ & $8.72 ^{+1.19}_{-1.25} $  & $7.5 _{-3.3}^{+ 6.0}$ & $0.99^{+0.66}_{-0.59}$  &$0.48  _{-0.47}   ^{+38.7}$        & 0.1     & --   &--       \\
 PG0953     & -25.6 & $151 ^{+22}_{-27}   $& $2.885\pm0.065 $& $8.26_{-0.09}^{+0.06}$ & $10.57^{+1.08}_{-1.11} $  & $2.3 _{-1.0}^{+ 1.8}$ & $0.63^{+0.51}_{-0.48}$  &$0.003 _{-0.0028} ^{+0.1}$   & 1.1     & --   & --      \\
 PG1211     & -24.0 & $101 ^{+23}_{-29}   $& $1.832\pm0.081 $& $7.61_{-0.15}^{+0.09}$ & $9.79 ^{+1.12}_{-1.19} $  & $3.0 _{-1.3}^{+ 2.3}$ & $0.45^{+0.59}_{-0.52}$  &$0.012 _{-0.01}   ^{+0.7}$     & 0.1     & --   &  --     \\
 PG1226     & -26.9 & $387 ^{+58}_{-50}   $& $3.416\pm0.072 $& $8.74_{-0.07}^{+0.07}$ & $11.53^{+1.08}_{-1.08} $  & $1.4 _{-0.6}^{+ 1.1}$ & $0.45^{+0.48}_{-0.48}$ &$2.11E-4_{-2.05E-4}^{+0.007}$   & 1621.2  & --   &   --    \\
 PG1229     & -22.4 & $50  ^{+24}_{-23}   $& $3.44 \pm0.12  $& $8.74_{-0.27}^{+0.17}$ & $8.85 ^{+1.21}_{-1.31} $  & $11.6_{-5.1}^{+ 9.6}$ & $0.44^{+0.75}_{-0.62}$  &$0.1   _{-0.097}  ^{+11.2}$        & 0.3     & --   &   --    \\
 PG1307     & -24.6 & $124 ^{+45}_{-80}   $& $4.19 \pm0.14  $& $8.45_{-0.45}^{+0.14}$ & $10.11^{+1.17}_{-1.55} $  & $5.2 _{-2.3}^{+ 4.8}$ & $0.49^{+0.96}_{-0.57}$  &$0.006 _{-0.006}  ^{+2.0}$    & 0.2     & --   &   --    \\
 PG1351     & -24.1 & $227 ^{+149}_{-72}  $& $1.17 \pm0.16  $& $7.66_{-0.23}^{+0.23}$ & $10.25^{+1.27}_{-1.21} $  & $1.6 _{-0.8}^{+ 1.3}$ & $0.05^{+0.61}_{-0.67}$  &$0.0016_{-0.0015} ^{+0.1}$   & 0       & --   &   --    \\
 PG1411     & -24.7 & $102 ^{+38}_{-37}   $& $2.456\pm0.096 $& $7.90_{-0.20}^{+0.14}$ & $10.04^{+1.17}_{-1.24} $  & $3.0 _{-1.3}^{+ 2.4}$ & $0.61^{+0.65}_{-0.58}$  &$0.01  _{-0.009}  ^{+0.7}$      & 0.2     & --   &   --    \\
 PG1426     & -23.4 & $95  ^{+31}_{-39}   $& $6.25 \pm0.39  $& $8.67_{-0.24}^{+0.13}$ & $9.55 ^{+1.15}_{-1.26} $  & $13.0_{-5.8}^{+10.5}$ & $0.34^{+0.70}_{-0.56}$  &$0.016 _{-0.015}  ^{+1.4}$     & 0.2     & --   &   --    \\
 PG1613     & -23.5 & $39  ^{+20}_{-14}   $& $7    \pm0.38  $& $8.38_{-0.20}^{+0.18}$ & $9.12 ^{+1.22}_{-1.21} $  & $15.4_{-6.8}^{+12.1}$ & $0.84^{+0.66}_{-0.63}$  &$0.13  _{-0.13}   ^{+10.0}$       & 0.8     & --   &   --    \\
 PG1617     & -23.4 & $85  ^{+19}_{-25}   $& $5.12 \pm0.85  $& $8.44_{-0.19}^{+0.12}$ & $9.49 ^{+1.11}_{-1.18} $  & $10.8_{-4.7}^{+ 8.4}$ & $0.40^{+0.60}_{-0.51}$  &$0.02  _{-0.02}   ^{+1.2} $        & 0.7     & --   &   --    \\
 PG1700$^{c}$     & -25.8 & $88  ^{+190}_{-182} $& $2.18 \pm0.17  $& $7.78_{-4.78}^{+0.51}$ & $10.36^{+1.62}_{-4.59} $  & $1.7 _{-0.8}^{+ 1.7}$ & $0.96^{+3.99}_{-1.01}$  &$0.01  _{-0.01}   ^{+3.9E6}$       & 8.5     & --   &   --    \\
 PG1704     & -25.6 & $319 ^{+184}_{-285} $& $0.89 \pm0.28  $& $7.57_{-4.57}^{+0.26}$ & $10.97^{+1.25}_{-2.19} $  & $0.6 _{-0.3}^{+0.81}$ & $0.23^{+1.59}_{-0.65}$ &$4.73E-4_{-4.6E-4} ^{+2.8}$    &562.8    & --   &   --    \\
 PG2130     & -22.9 & $200 ^{+67}_{-18}   $& $2.41 \pm0.15  $& $8.16_{-0.05}^{+0.13}$ & $9.76 ^{+1.16}_{-1.05} $  & $5.7 _{-2.5}^{+4.20}$ &$-0.18^{+0.46}_{-0.56}$  &$0.003 _{-0.003}  ^{+0.1}$     &0.4      & --    &  --    \\

\hline
\end{tabular}

\caption{The properties of the 34 AGNs. Col.1: name, Col.2:
absolute B band magnitude from Veron-Cetty et al. (2001), Col.3:
size of BLRs in lt days from Kaspi et al. (2001), Col.4: FWHM (in
$1000km s^{-1}$) of H$\beta$ from Kaspi et al. (2000), Col.5:log
of the reverberation mapping BH mass in $\Msolar$ from Kaspi et
al. (2000), Col.6:log of our calculated BH mass in $\Msolar$ ,
Col.7: calculated inclinations (in deg) to our sight, Col.8: log
of accretion rates in $\Msolar/yr$ , Col.9 the accretion rate in
units of Eddington accretion rate , Col.10: the radio loudness
from Nelson (2001) , Col.11, the nuclear radio loudness from Ho,
L. C. and Peng, C. Y. (2001), Col.12, inclinations from Wu \& Han
(2001). $^{a},^{b},^{c}$: the lower limit of the size of BLRs is
not larger than zero and it leads to the very high accretion
rate.}
\end{tiny}
\end{center}

\end{table*}

\begin{figure}
\centerline{\epsfig{file=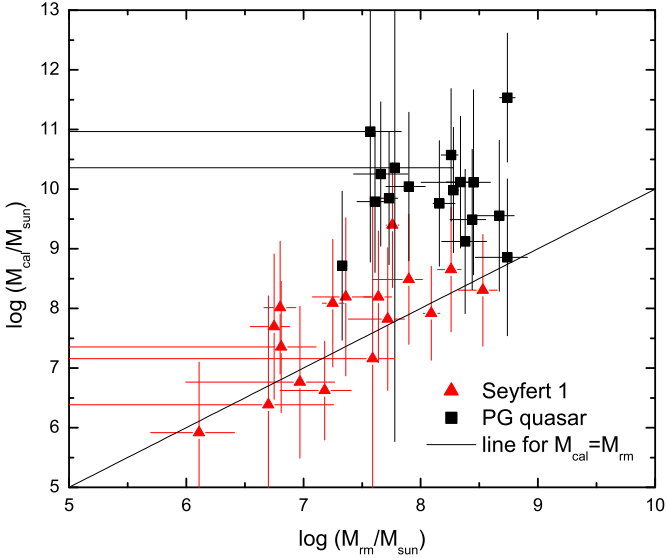,width=10cm,angle=0,clip=}}
\caption{$M_{cal}$ versus $M_{rm}$($\alpha=1$). The triangle
denotes Seyfert 1 galaxy and the square denotes PG quasar. The
beeline represents $M_{cal}=M_{rm}$.}
\end{figure}

\subsection{Masses}

Assuming random disk inclinations to the line of sight,  these 34
central black hole reverberation mapping masses $M_{rm}$ have been
obtained (Kaspi et al. 2000). Here we plot in Fig.1 our calculated
central black hole masses considering the effects of inclinations
($M_{cal}$) versus the reverberation mapping masses ($M_{rm}$).
Fig.1 shows that, though $M_{cal}$ of 17 PG quasars are lager than
their $M_{rm}$ , $M_{cal}$ of 17 Seyfert 1 galaxies are consistent
with $M_{rm}$ of them. We have a $\chi^{2}$ (Press et al. 1992,
p616) test to show if our calculated BH masses $M_{cal}$ are
consistent with $M_{rm}$ from Kaspi et al. (2000). $\chi^{2}$ and
probability are 1.81 and 99.9\% for 17 Seyfert 1 galaxies. For 17
PG quasars $\chi^{2}$ and probability are  10.98 and 81\%
respectively.

\subsection{Inclinations}

By fitting the observed iron $K\alpha$ line profile with the
accretion disk model, Nandra et al. (1997) estimated the
inclinations, with a mean value of $30^{o}$, for 18 Seyfert 1
galaxies. Wu \& Han (2001) developed a simple method to drive the
inclinations (with a mean value of $36^{o}$) for a sample of 11
Seyfert 1 galaxies that have both measured bulge velocity
dispersion and BH masses estimated by reverberating mapping
technique. The mean and the rms of the mean of our calculated
inclinations for 17 Seyfert 1 galaxies are $32.2\pm5.5$ (deg),
which are consistent with the results of Nandra et al. ( 1997 )
and Wu \& Han (2001). In fig.2 we plot inclinations from Wu \& Han
(2001) versus our calculated inclinations for 9 common Seyfert 1
galaxies. The dash line in Fig.2 presents inclinations from Wu \&
Han (2001) and our calculated inclinations are consistent. Our
calculated inclinations of 17 Seyfert 1 galaxies provide further
support to the orientation-dependent unification scheme of AGNs.
The mean and error of our calculated inclinations for 17 PG
quasars are $5.65\pm1.08$ (deg), which are smaller compared to
that of Seyfert 1 galaxies.

\begin{figure}
\centerline{\epsfig{file=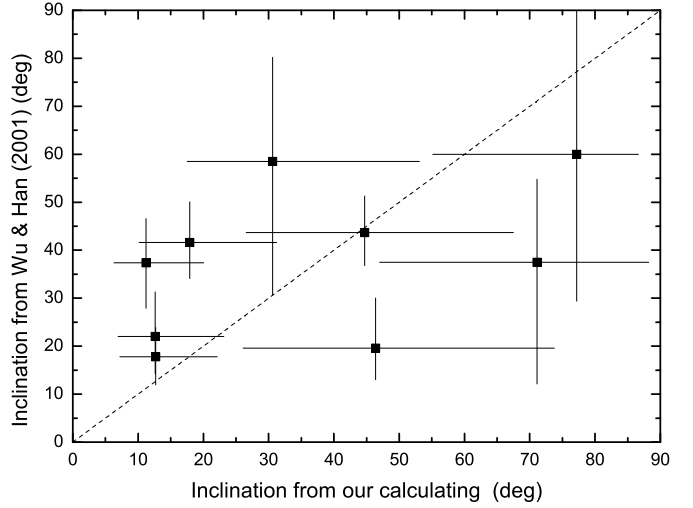,width=10cm,angle=0,clip=}}
\caption{Inclinations form Wu \& Han (2001) versus our calculated
inclinations. The dash line presents inclinations from Wu \& Han
(2001) and our calculated inclinations are consistent. }
\end{figure}

A positive correlation of inclinations with observed FWHMs of the
$H\beta$ line has been found (Wu \& Han 2001). Fig. 3 displays the
inclination of the disk to the line of sight versus the width of
H$\beta$. We can find there is a correlation for Seyfert 1
galaxies between FWHMs of H$\beta$ and inclinations. A simple
least square linear regression (Press et al. 1992, p655) gives
$i=(12.95\pm10.65)+(4.7\pm2.29)(V_{FWHM}/1000\rm{km s^{-1}})$,
with a Pearson correlation coefficient of R=0.47 corresponding to
a probability of P=0.058 that the correlation is caused by a
random factor. A minimum $\chi^{2}$ fit considering the errors of
both parameters ( Press et al. 1992, p660 ) gives
$i=(-0.80\pm4.43)+(4.15\pm1.59)(V_{FWHM}/1000\rm{km s^{-1}})$ ,
with $\chi^{2}$ and probability of 6.04 and 73\%, respectively.
There also exists a correlation between FWHMs of H$\beta$ and
inclinations for 17 PG quasars. A simple least square linear
regression gives
$i=(-1.75\pm1.3)+(2.29\pm0.36)(V_{FWHM}/1000\rm{km s^{-1}})$, with
R=0.85 (P$<10^{-4}$). A minimum $\chi^{2}$ fit considering the
errors of both parameters gives
$i=(1.85\pm1.47)+(0.19\pm0.53)(V_{FWHM}/1000\rm{km s^{-1}})$ ,
with $\chi^{2}$ and probability of 6.9 and 65\%, respectively.

\begin{figure}
\centerline{\epsfig{file=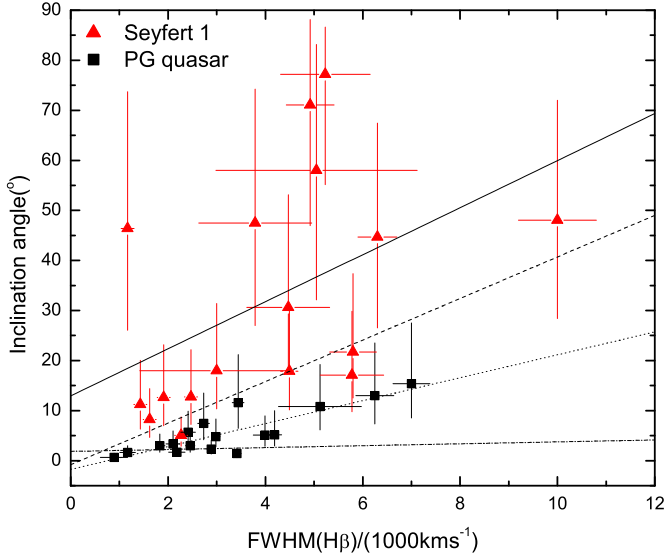,width=10cm,angle=0,clip=}}
\caption{Inclination versus width of H$\beta$($\alpha=1$). The
denotation is the same as that in Fig.1. The solid line and dash
line show the simple least square linear fit for Seyfert 1
galaxies and Quasars, respectively. The dot line and dash dot line
show the minimum $\chi^{2}$ fit with errors of both parameters
considered for Seyfert 1 galaxies and quasars, respectively. }
\end{figure}

A possible anticorrelation of inclinations with the nuclear radio
loudness has been found (Wu \& Han 2001). Fig.4 displays the
inclinations versus the radio loudness, defined by the ratio
between 5 GHz radio luminosity and B-band optical luminosity
(Nelson 2001). Because there exist the contaminations to the
luminosity of galactic nucleus from the host galaxies, using the
nuclear radio loudness will be better to describe the nature of
galactic nuclei. We plot in Fig.5 the inclinations versus the
nuclear radio loudness, defined by the ratio between 5 GHz nuclear
radio luminosity and B-band nuclear optical luminosity (Ho \& Peng
2001). Although there are only eight AGNs with available data for
nuclear radio loudness, Fig.5 still shows the tendency that
Seyfert 1 galaxies with the larger nuclear radio loudness may have
smaller inclinations. A simple least square linear regression
gives $i=(53.9\pm17.9)+(-13.0\pm11.8)(R_{nuc})$, with a
correlation coefficient of R=0.41 corresponding to a probability
of P=0.31 that the correlation is caused by a random factor. It is
consistent with the result of Wu \& Han (2001).

\begin{figure}
\centerline{\epsfig{file=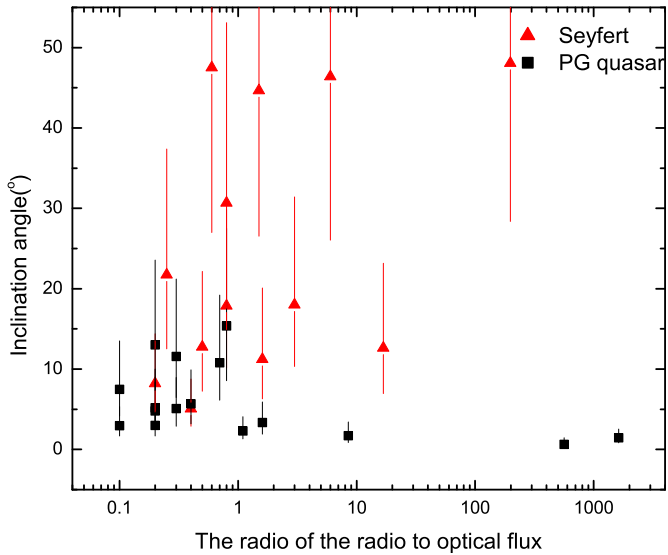,width=10cm,angle=0,clip=}}
\caption{Inclinations versus the radio loudness($\alpha=1$). The
denotation is the same as that in Fig.1.}
\end{figure}

\begin{figure}
\centerline{\epsfig{file=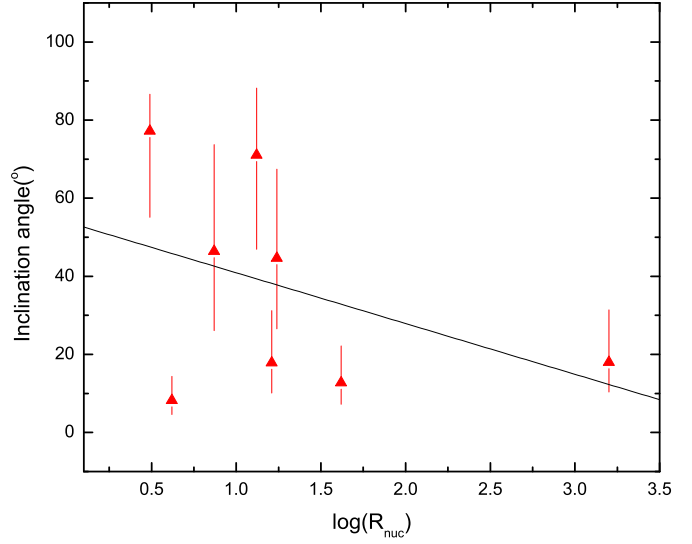,width=10cm,angle=0,clip=}}
\caption{Inclinations versus the nuclear radio
loudness($\alpha=1$). The denotation is the same as that in Fig.1.
The solid line shows the simple least square linear fit. }
\end{figure}

Considering the disk inclinations we calculate the intrinsic width
of H$\beta$, which is approximately equal to the observed width of
H$\beta$ divided by $sin(i)$. In Fig.6 we plot the intrinsic width
of H$\beta$ versus the observed width of H$\beta$.
\begin{figure}
\centerline{\epsfig{file=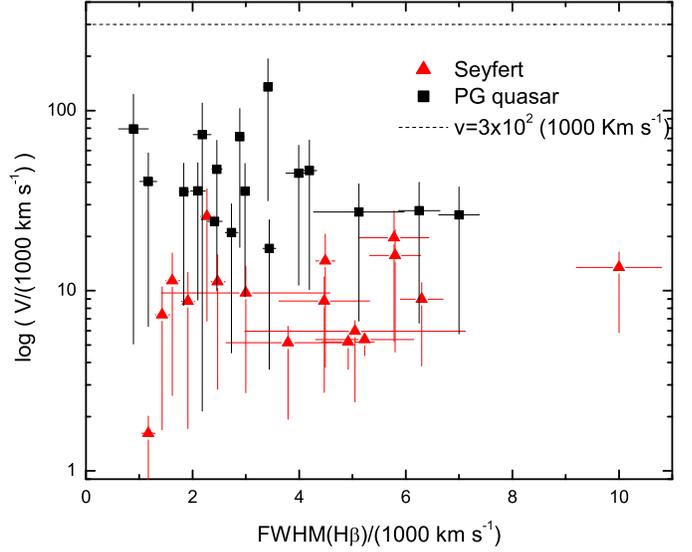,width=10cm,angle=0,clip=}}
\caption{The intrinsic width of H$\beta$ versus the observed width
of H$\beta$($\alpha=1$). The straight dash line shows the velocity
of light. The denotation is the same as that in Fig.1.}
\end{figure}

\subsection{Accretion Rates}

In Fig.7 we plot $M_{cal}$ versus the accretion rates $\dot{M}$
and in Fig.8 we plot $M_{cal}$ versus the accretion rates in terms
of the Eddington accretion rate. Our calculated accretion rates
for PG quasars are consistent with that from Laor (1990). Our
calculated accretion rates for Seyfert 1 galaxies are larger and
we discuss it in next section.

\begin{figure}
\centerline{\epsfig{file=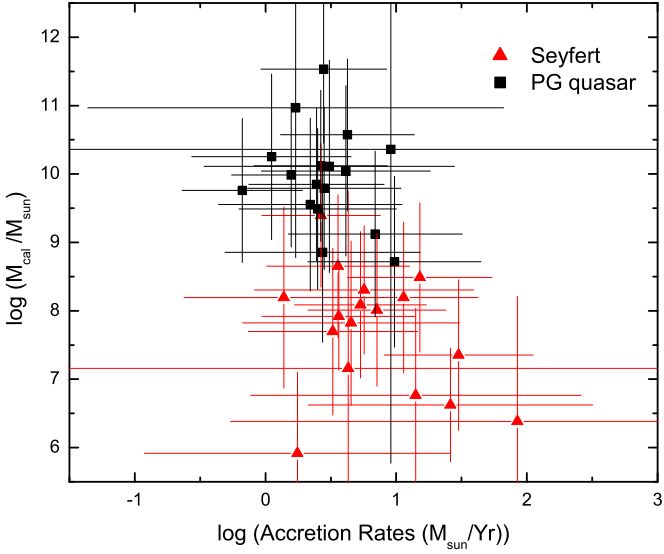,width=10cm,angle=0,clip=}}
\caption{$M_{cal}$ versus the accretion rates (\Msolar/yr)
($\alpha=1$). The denotation is the same as that in Fig.1.}

\end{figure}

\begin{figure}
\centerline{\epsfig{file=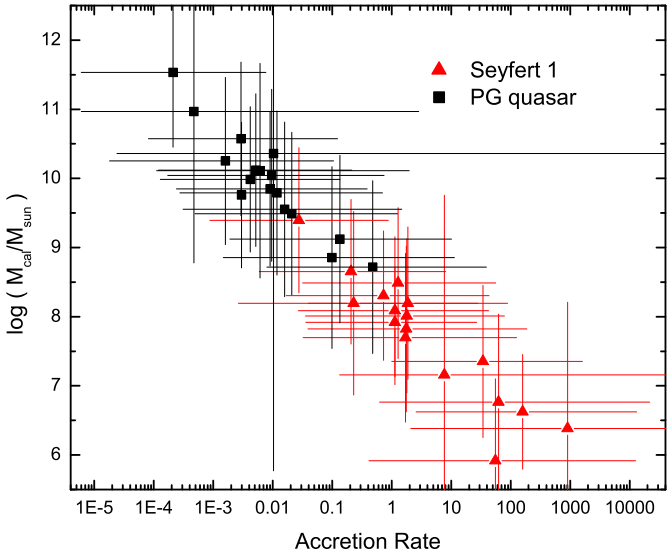,width=10cm,angle=0,clip=}}
\caption{$M_{cal}$ versus the accretion rate in units of the
Eddington accretion rate($\alpha=1$). The denotation is the same
as that in Fig.1.}

\end{figure}

\subsection{$M_{bh}-\sigma$ Relation}

Ferrarese et al. (2001) measured bulge velocity dispersions of
AGNs and found good agreement between BH masses obtained from
reverberating mapping technique and from the $M_{bh}-\sigma$
relation as defined by quiescent galaxies. In Fig.9 we plot our
calculated BH masses considering inclinations versus the bulge
velocity dispersion of AGNs. The dash line in Fig.9 is the best
linear fit to the $M_{bh}-\sigma$ relation as published by Merritt
\& Ferrarese (2001) for nearby normal galaxies. Although there is
a large scatter our calculated BH mass is also consistent with the
$M_{bh}-\sigma$ relation as defined by quiescent galaxies. A
minimum $\chi^{2}$ fit for our calculated BH mass and the bulge
velocity dispersion considering error of  both parameters gives
$log(M/(10^{7}\Msolar))=(-1.63\pm5.49)+(4.32\pm2.65)log(\sigma/(\rm{km
s^{-1}))} $, with $\chi^{2}$ and probability of 1.55 and 98\%,
respectively. The minimum $\chi^{2}$ fit for the reverberation
mapping BH mass (Kaspi et al. 2000) and  the bulge velocity
dispersion gives
$log(M/(10^{7}\Msolar))=(-1.64\pm1.17)+(4.3\pm0.55)log(\sigma/(\rm{km
s^{-1}))}$ ($\chi^{2}=14.2$ and probability is 0.05), which is
very similar to our results.

\begin{figure}
\centerline{\epsfig{file=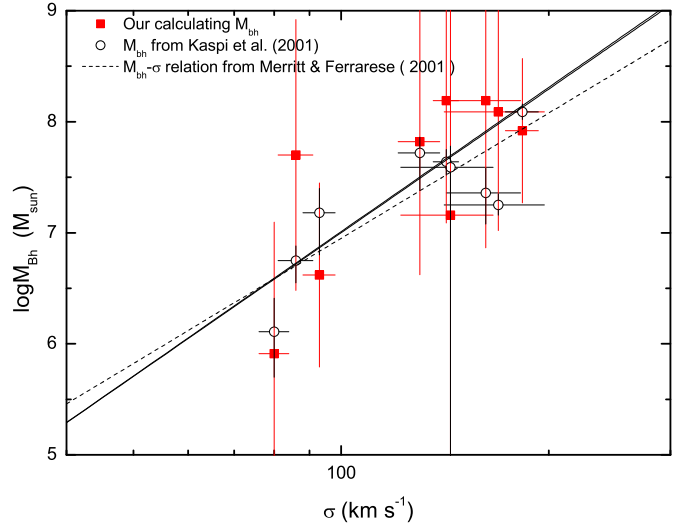,width=10cm,angle=0,clip=}}
\caption{Black hole mass versus velocity dispersion in the bulges
of the hosts of AGNs. Solid square  denotes our calculated BH mass
and open circle denotes the reverberating mapping BH mass. The
dash line is the best linear fit to the $M_{bh}-\sigma$ relation
as published by Merritt \& Ferrarese (2001) for nearby normal
galaxies. The solid line is a minimum $\chi^{2}$ fit line for our
calculated BH mass and bulge velocity dispersion considering the
errors of both parameters.}

\end{figure}

\subsection{Uncertainties of Our Results}
We also adopt $\alpha=0.1$ and $\alpha=0.01$ to calculate $\Delta
(logM)=log(M_{cal})-log(M_{rm})$ (Q=1 and not considering the
uncertainties of sizes of BLRs and absolute B magnitudes). The
mean and the standard error(SE) of $\Delta (logM)$ are calculated.
The results are shown in Table2. From Table2, we can find the mean
of $\Delta log(M)$ for Seyfert 1 galaxies is smaller than that of
PG quasars with the same value of $\alpha$. With the decrease of
$\alpha$ the mean of $\Delta log(M)$ becomes larger. In table 3,4
we list the mean and error of uncertainties of our results for
Seyfert 1 galaxies and PG quasars, respectively. Considering the
effect of uncertainty of B band magnitude we find the
uncertainties of calculated inclinations, masses and accretion
rates will be larger. The errors of BLRs sizes don't effect our
calculated inclinations too much compared to the influence of B
magnitude.

\begin{table}
\begin{center}
\begin{tabular}{lccc}
\hline \hline

$\alpha$ &1&0.1&0.01\\
\hline
Seyfert 1 & 0.33$\pm$0.15&0.99$\pm$0.18&1.72$\pm$0.19\\
PG quasar& 1.83$\pm$0.20&2.58$\pm$0.20&3.34$\pm$0.20\\
\hline
\end{tabular}
\caption{The mean and the error of $\Delta M$ in logarithm  for
Seyfert 1 galaxies and PG quasars considering $\alpha=1, 0.1,
0.01$ and $Q=1$.}
\end{center}
\end{table}

\begin{table*}
\begin{center}

\begin{tabular}{lccccccccc}
\hline \hline

& $\delta i_{1}$ &$\delta i_{2}$&$\delta i_{3}$&$\delta
\dot{M}_{1}$&$\delta \dot{M}_{2}$&$\delta \dot{M}_{3}$& $\delta
M_{1} $&$\delta M_{2}$& $\delta M_{3}$\\
&(1)&(2)&(3)&(4)&(5)&(6)&(7)&(8)&(9)\\
\hline
upper limit &15.10 $\pm$ 1.98&13.78$\pm$1.72&3.17$\pm$0.46&0.73$\pm$0.26&0.49$\pm$0.09&0.67$\pm$0.02&1.10$\pm$0.03&0.95$\pm$0.02&0.63$\pm$0.007\\
\hline
lower limit & 11.16$\pm$7.06&10.56$\pm$6.47&2.91$\pm$1.65&0.60$\pm$0.14&0.45$\pm$0.06&0.66$\pm$0.02&1.09$\pm$0.04&0.89$\pm$0.04&0.63$\pm$ 0.007\\

\hline
\end{tabular}
\caption{The mean and the error of uncertainties of our results
for 15 Seyfert 1 galaxies (exclude IC4329A, NGC3227 for their
unavailable lower limit of BLRs sizes). Column (1) to (3) are
uncertainties of calculated inclinations, $\delta i$. Column (4)
to (6) are uncertainties of calculated accretion rates, $\delta
\dot{M}$. Column (7) to (9) are uncertainties of calculated BH
masses, $\delta M$. Column (3), (6) and (9) is the values just
considering the effect of uncertainty of Q. Column (2), (5) and
(8) is the values considering the effect of uncertainties of Q and
B magnitude. Column (1), (4) and (7) is the values considering the
effect of uncertainties of Q, B magnitude and size of BLRs.}

\end{center}
\end{table*}

\begin{table*}
\begin{center}

\begin{tabular}{lccccccccc}
\hline \hline

& $\delta i_{1}$ &$\delta i_{2}$&$\delta i_{3}$&$\delta
\dot{M}_{1}$&$\delta \dot{M}_{2}$&$\delta \dot{M}_{3}$& $\delta
M_{1} $&$\delta M_{2}$& $\delta M_{3}$\\
&(1)&(2)&(3)&(4)&(5)&(6)&(7)&(8)&(9)\\
\hline
upper limit &4.72 $\pm$ 0.90&4.23$\pm$0.80&0.84$\pm$0.16&0.67$\pm$0.07&0.41$\pm$0.001&0.65$\pm$0.001&1.15$\pm$0.02&1.00$\pm$0.001&0.65$\pm$0.002\\
\hline
lower limit & 2.58$\pm$0.49&2.47$\pm$0.47&0.73$\pm$0.14&0.56$\pm$0.02&0.41$\pm$0.002&0.65$\pm$0.001&1.26$\pm$0.07&1.00$\pm$0.002&0.65$\pm$ 0.003\\

\hline
\end{tabular}
\caption{The mean and the error of uncertainties of our results
for 16 PG quasars (exclude PG1700 for unavailable lower limit of
BLRs sizes). Column (1) to (3) are uncertainties of calculated
inclinations, $\delta i$. Column (4) to (6) are uncertainties of
calculated accretion rates, $\delta \dot{M}$. Column (7) to (9)
are uncertainties of calculated BH masses, $\delta M$. Column (3),
(6) and (9) is the values just considering the effect of
uncertainty of Q. Column (2), (5) and (8) is the values
considering the effect of uncertainties of Q and B magnitude.
Column (1), (4) and (7) is the values considering the effect of
uncertainties of Q, B magnitude and size of BLRs.}

\end{center}
\end{table*}

\section{Discussions and Conclusions}

\subsection{The large accretion rates for Seyfert 1 galaxies.}

In Fig.7 and Fig.8, though the accretion rates in units of the
Eddington accretion rates for PG quasars are consistent with the
results of Laor (1990), we can see that accretion rates in units
of the Eddington accretion rates of  the most of Seyfert 1
galaxies is larger than one. We confirmed the results of Collin \&
Hure (2001). They also find it radiates at super-Eddington rates
if a standard accretion disc accounts for the observed optical
luminosity. They can't account for the effect of the inclination.
In this paper we consider the effect of the inclinations and
uncertainties of accretion rates from Q, BLRs sizes and B absolute
magnitude. Although we find there also exists high accretion rates
for Seyfert 1 galaxies, we should notice the uncertainties about
accretion rates (Fig. 8). The large accretion rates are maybe due
to the uncertainty of dynamics of BLRs, median absolute B
magnitudes. More knowledge of BLRs dynamics, accretion disks and
optical luminosity are needed to improve the determinations of
accretion rates in AGNs.

\subsection{The disk inclination to the line of sight}

With our calculation, there is apparent difference in inclinations
between Seyfert 1 galaxies and PG quasars (Fig.3, Table 1) when we
adopt  $\alpha=1$ for all objects. Inclinations for quasars are
smaller than that for Seyfert galaxies. The mean and error of
inclinations for 17 Seyfert 1 galaxies are $32.2\pm5.5$ (deg)
($\alpha=1$). The mean and error of inclinations for 17 PG quasars
are $5.65\pm1.08$ (deg) ($\alpha=1$). Only 4 quasars have
inclinations larger than $10^{o}$. The mean uncertainty of
inclinations of quasars is large, about $4.7^{o}$ (Table 4). The
mean value of inclinations for PG quasar can be about $10^{o}$ as
upper limit considering the uncertainties of Q, BLR sizes and B
magnitude. We noticed that the ratio between the black hole mass
determined by reverberation mapping technique ($M_{rm}$) and our
calculated black hole mass ($M_{cal}$) can be approximated by
$3(sini)^{2}$ (Wu \& Han, 2001). With the mean value of
inclinations ($32^{o}$) derived by us, we obtained
$M_{rm}/M_{cal}=1.19$ for Seyfert 1 galaxies. It means that the
black hole mass estimated by the standard method of reverberation
mapping can still represent the "true" black hole mass well if the
inclination of the Seyfert galaxy is not substantially different
from $32^{o}$. With the mean value of inclinations ($5.65^{o}$)
derived by us, we obtained $M_{rm}/M_{cal}=34$ for PG quasars.
According to our model and when using $\alpha=1$, it is necessary
to consider the effect of inclinations to calculate the central
black hole masses for quasars.

Based on AGNs unification schemes, AGNs with wide emission lines
from BLRs are oriented at a preferred angle from which BLRs is
visible. No edge-on disks would be seen (Urry \& Padovani 1995).
Our calculated inclinations, with a mean value of $32^{o}$ for 17
Seyfert 1 galaxies that agrees well with the result obtained by
fitting the iron $K\alpha$ lines of Seyfert 1 galaxies observed
with ASCA (Nandra et al. 1997) and the result obtained by Wu \&
Han (2001), provide further support for the orientation-dependent
unification scheme of active galactic nuclei. Based on the unified
model of AGNs, the wide variety of AGN phenomena we see is due to
a combination of real differences in a small number of physical
parameters (e.g. luminosity) coupled with apparent differences
which are due to observer-dependent parameters (e.g. orientation).
The unification scheme does not explain the difference between
Seyfert 1 galaxies and quasars in an inclination effect (as they
both should have the same inclinations of close to face on) but
explain it as a luminosity difference. We here find that
inclinations of PG quasars are smaller than that of Seyfert 1
galaxies if we adopt the same value of $\alpha$ for all objects (
Fig.3, Table 1) . Do luminous quasars prefer to face on ? We adopt
different values of parameter $\alpha$ and try to bring quasars
inclinations to the inclination levels of Seyfert 1 galaxies. We
use larger $\alpha$ to recalculated inclinations for PG quasars.
The mean and error of inclinations for 17 PG quasars is
$13.4^{+10.4}_{-5.8}$ (deg) ($\alpha=10$) and
$29.7^{+15.0}_{-12.0}$ (deg) ($\alpha=100$), which are in the
inclination levels of Seyfert 1 galaxies. The errors of Q,
absolute B band magnitude, BLRs sizes are considered to calculate
the uncertainties of inclinations. It is possible that the
difference between PG quasars and Seyfert 1 galaxies is maybe due
to the difference of the value of $\alpha$. From above all we may
find calculated inclinations of PG quasars are smaller than that
of Seyfert 1 galaxies unless we adopt larger value of $\alpha$ for
PG quasars to calculate in our model. However we often assume that
the value of $\alpha$ of disc in AGNs is often between 0 and 1
(Shakura \& Sunyaev 1973) . The need of higher value of $\alpha$
(larger than one) for PG quasars maybe shows that our model is not
suitable for PG quasars if we think inclinations of PG quasars are
in the inclination levels of Seyfert 1 galaxies.

In Fig.3 we can find the NLS1s (here we simply define the NLS1s
with FWHM of H$\beta$ is less than 2000 $km s^{-1}$) have the
smaller inclinations except for NGC4051. We should notice the
larger scatter about the value of inclination for NGC4051 (Table
1). Collin \& Hure (2001) suggested the sizes of the BLRs will
increase with the accretion rates expressed in Eddington units and
decrease with the black hole masses. The larger size of BLRs led
to the smaller widths of H$\beta$ in NLS1s. They omitted the
effect of the inclinations. The virial BH mass is given by
$M=\frac{V_{FWHM}^{2}}{4(sini^{2}+A^{2})}RG^{-1}$. Because the
difference for the sizes of BLRs is not big, NLS1s with small
values of $V_{HWHM}$ will have smaller BH masses when we don't
consider the effect of inclinations. The effect of inclinations in
NLS1s should be considered when we study the physics of NLS1s. In
Fig.6 we show the velocity of BLRs and we find the intrinsic
widths of H$\beta$ for NLS1s are not smaller than that of broad
line AGNs, which is different from the results of Wu \& Han
(2001), which is from 11 Seyfert 1 galaxies.

\subsection{The size of BLRs}

There is a natural idea that BLRs are made of the atmosphere or
winds of giant stars (Edwards 1980). Another idea is that BLRs are
from the accretion disk or the wind released at the top of the
disk (Murray \& Chiang 1997). With the reasonable values of
$M_{cal}$ and inclinations, we show that the gravitational
instability can indeed explain the size of BLRs. Assuming the
gravitational instability of standard thin accretion disk leads to
the Broad Line Regions(BLRs), the B band luminosity comes from
standard thin disk and the motion of BLRs is virial, we can obtain
$R_{BLR} \propto L_{B}^{0.5}\dot{M}^{-37/45}$ from Eq. 1 and Eq.
2. If $L_{5100} \propto L_{B}$, there is a correlation $R_{BLR}
\propto L_{5100}^{0.5}\dot{M}^{-37/45}$. The size of BLRs relates
not only to the luminosity, but also to the accretion rates.
Nicastro (2000) proposed that the BLRs are released by the
accretion disk in the region where a vertically outflowing corona
exists and he found the sizes of BLRs would relate to $\dot{M}$.
Collin \& Hure (2001) showed the size of BLRs mainly related to
$M$, not $\dot{M}$. Based on the formulae $R_{BLR} \propto
L_{5100}^{0.5}\dot{M}^{-37/45}$, the difference of the accretion
rate for the Seyfert 1 galaxies is not much and there is a
relation as $R_{BLR} \propto L_{5100}^{0.5}$, the power index is
almost 0.5, which can explain the correlation between the size of
BLRs($R_{BLR}$) and the monochromatic luminosity at 5100\AA
($L_{5100}$), $R_{BLR} \propto L_{5100}^{0.5}$ (Wandel et al.
1999). The $\dot{M}$ is different for Seyfert 1 galaxies and
quasars. The $\dot{M}$ of luminous quasars are small (see Fig.7,
Fig.8). The index will be higher for the sample of Kaspi et al.
(2000) which includes 17 Seyfert 1 galaxies and 17 PG quasars. It
is necessary to check this with a larger sample.

\subsection{Conclusion}

With the formulae of the standard thin disk and the assumption of
gravitational instability leading to BLRs, we calculate the
central black hole masses, accretion rates and inclinations for 34
AGNs. The main conclusions can be summarized as follows:
\begin{itemize}
\item{The gravitational instability can indeed explain the size of BLRs.
The size of BLRs relates not only to the luminosity, but also to the
accretion rates.}
\item{Our results are sensitive to $\alpha$ parameter of the standard
$\alpha$ disk. $\alpha$ is 1 in all the calculations. The mean
value of inclinations for 17 Seyfert 1 galaxies is $32^{o}$ ,
which is favoring the orientation-dependent unification scheme of
AGNs. Inclinations of 17 Seyfert 1 galaxies are about 6 times
larger than that of 17 PG quasars unless the value of $\alpha$ of
PG quasars is larger than Seyfert 1 galaxies. The need of higher
value of $\alpha$ for PG quasars maybe shows that our model is not
suitable for PG quasars if we think inclinations of PG quasars are
in the inclination levels of Seyfert 1 galaxies. There is a
correlation between inclinations and the FWHMs of H$\beta$. Though
the observed FWHMs of H$\beta$ for NLS1s is smaller than that for
broad line AGNs, the intrinsic of FWHMs of H$\beta$ for NLS1s are
not smaller than that of broad line AGNs. Small inclinations lead
to the small FWHMs of H$\beta$ for NLS1s.}
\item{The uncertainty of absolute B band magnitude (here we adopt
1 magnitude) from variability or the hosts contribution leads to
larger scatter of our results than error of size of BLRs. More
knowledge of BLRs dynamics, accretion disks and optical luminosity
are needed to improve the determinations of black hole masses,
accretion rates and inclinations in AGNs.}

\end{itemize}

\acknowledgements

We thank Keliang Huang for useful discussions, and the anonymous
referee for the valuable comments.

\end{document}